# The Multi-way Relay Channel

Deniz Gündüz*, Aylin Yener†, Andrea Goldsmith‡, H. Vincent Poor§

*CTTC, Barcelona, Spain

‡Dept. of Electrical Eng., Pennsylvania State Univ., University Park, PA

‡Dept. of Electrical Eng., Stanford Univ., Stanford, CA

§Dept. of Electrical Eng., Princeton Univ., Princeton, NJ

Email: deniz.gunduz@cttc.es, yener@ee.psu.edu, andrea@wsl.stanford.edu,

poor@princeton.edu

### Abstract

The multiuser communication channel, in which multiple users exchange information with the help of a relay terminal, termed the *multi-way relay channel* (mRC), is introduced. In this model, multiple interfering clusters of users communicate simultaneously, where the users within the same cluster wish to exchange messages among themselves. It is assumed that the users cannot receive each other's signals directly, and hence the relay terminal in this model is the enabler of communication. In particular, *restricted encoders*, which ignore the received channel output and use only the corresponding messages for generating the channel input, are considered. Achievable rate regions and an outer bound are characterized for the Gaussian mRC, and their comparison is presented in terms of exchange rates in a symmetric Gaussian network scenario. It is shown that the compress-and-forward (CF) protocol achieves exchange rates within a constant bit offset of the exchange capacity independent of the power constraints of the terminals in the network. A finite bit gap between the exchange rates achieved by the CF and the amplify-and-forward (AF) protocols is also shown. The two special cases of the mRC, the *full data exchange* model, in which every user wants to receive messages of all other users, and the *pairwise data exchange* model which consists of multiple two-way relay channels, are investigated in This work was presented in part at the 2009 IEEE International Symposium of Information Theory in Seoul, South Korea.

This research was supported by the National Science Foundation under Grants CNS-09-05398, CNS-07-16325, CNS-07-21445, CCR-02-37727, CNS-09-05086 the DARPA ITMANET program under Grant 1105741-1-TFIND and Grant W911NF-07-1-0028, and the U.S. Army Research Office under MURI award W911NF-05-1-0246.

detail. In particular for the pairwise data exchange model, in addition to the proposed random coding based achievable schemes, a nested lattice coding based scheme is also presented and is shown to achieve exchange rates within a constant bit gap of the exchange capacity.

## I. INTRODUCTION

Relay terminals in wireless networks are instrumental in providing robustness against channel variations, extending coverage in the case of power limited terminals, and in improving energy efficiency. The three-terminal relay channel [1], one of the earliest models in network information theory, serves as a main building block for large wireless networks. More recently, it has been recognized that effective relaying protocols can be devised to facilitate cooperation between two users when they want to exchange information simultaneously over a single relay terminal. In this paper we introduce a new fundamental building block for general multicast communication. The model, termed the *multi-way relay channel (mRC)*, considers multiple clusters of users such that the users in each cluster want to exchange information among themselves. This exchange is facilitated by a relay terminal that helps all the users in the system. We consider a total of $N$ users grouped into $L \geq 1$ clusters of $K \geq 2$ distinct users each, i.e., $N = KL$.

This setup is general enough to model a variety of communication scenarios. Consider, for example, a peer-to-peer wireless network with groups of users sharing data with the help of a relay node. Here, the users who are interested in the same file can be grouped into clusters. Each user in the cluster might have a portion of the file that is required by the other users in the cluster. Many such clusters need to be served simultaneously by the relay terminal. Similarly, in a social network scenario, the clusters may be formed based on the connections among the users, and the users in each cluster, or a friend group, might want to exchange their personal information through the relay terminal (see Fig. 1 for an illustration). In a sensor network scenario, clusters may be formed based on the physical phenomenon that the sensors are measuring, i.e., temperature sensors exchange local temperature among themselves while pressure sensors exchange local pressure measurements. As yet another example, consider multiple terrestrial (ad-hoc) networks with nodes geographically distributed and served by a single communication satellite. The nodes in each network may want to exchange available local information (e.g. control information) among all the network nodes.

Our focus for this new model will be on the Gaussian channel. In particular, we have an additive Gaussian multiple access channel (MAC) from the users to the relay terminal, and a Gaussian broadcast channel from the relay to the users. It is assumed that the users do not receive each other's signals. This can be due to physical restrictions among the nodes in the sensor network scenario, due to large physical distances among the terminals in the satellite scenario, or due to the network protocol since infrastructure-based systems tend to have better performance and lower latency than ad-hoc ones. We study the most fundamental coding techniques that have been introduced in the literature for relay networks. In particular, we derive the achievable rate regions for the corresponding multi-way extensions of decode-and-forward (DF), amplify-and-forward (AF) and compress-and-forward (CF) protocols.

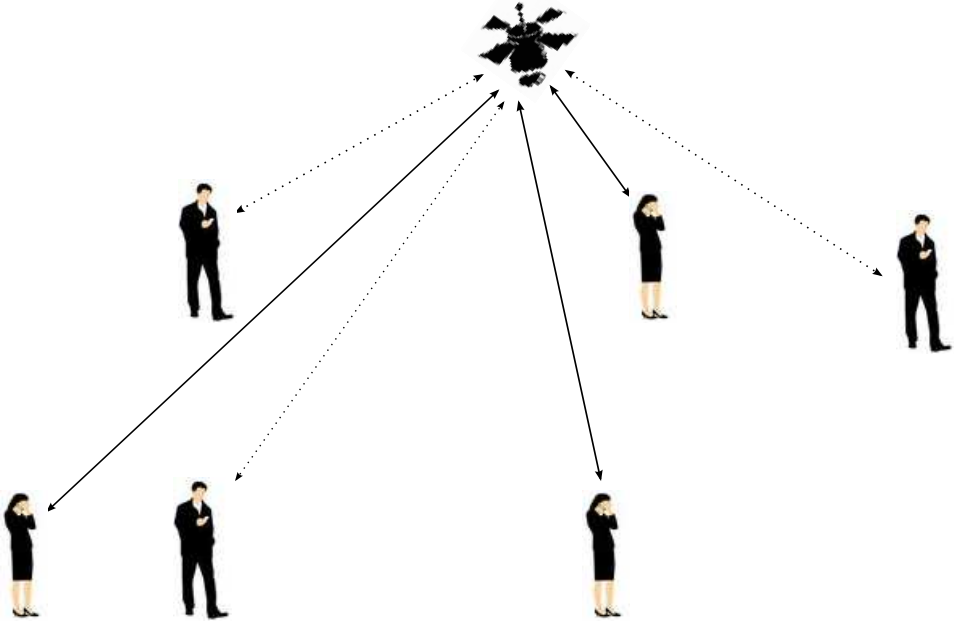

Fig. 1. An illustration of the multi-way relay channel model in which the relay terminal helps two separate clusters of users in a social network to simultaneously exchange messages.

In the DF scheme, the relay node is enforced to decode all the messages in the network. Since relay node is not originally a sink node, the decoding at the relay is not imposed by the problem, and hence, can be limiting with the increasing number of users in the system. In the CF scheme the relay quantizes its observation and broadcasts the quantized version to all the receivers. This circumvents the decoding requirement at the relay, but it suffers from the fact that the noise at the relay terminal is also quantized and forwarded to the users.

In all these achievable schemes, we exploit the fact that each user already knows its own message and can use this information for more effective decoding. While for the AF scheme, this can be achieved by each user simply subtracting its own transmit signal from the signal it receives; in the case of DF and CF schemes, this requires using a more involved coding scheme introduced in [2] and [3], respectively, for lossless and lossy broadcasting of a common source to receivers with correlated side information.

To provide a performance comparison of the proposed coding schemes, we focus on the achievable symmetric rate, termed the *exchange rate*, for a symmetric network setup. We define the total exchange rate as the total rate of data that can be transmitted over the system while each user's message has the same rate. The supremum of achievable total exchange rates is called the exchange capacity. The investigation of the exchange capacity allows us to obtain simple explicit rate expressions, and acquire fundamental insights into the behavior of the communication protocols in consideration. We characterize analytically the exchange capacity upper bound and the total exchange rates achievable by AF, DF and CF schemes.

We investigate two special cases of the mRC in detail: the *full data exchange* model in which each user wants to learn all the messages in the network, and the *pairwise data exchange* model in which multiple user pairs exchange information. It is shown in [4] that the CF scheme achieves within a half bit of the capacity for the symmetric TRC, while DF achieves the capacity when the additional sum-rate constraint is not the bottleneck. Similarly, nested lattice codes [5] are shown in [6] to achieve rates within a half bit of the capacity in a TRC. Here, we show that similar finite-bit approximation results for the exchange capacity can be obtained in the more general model of the mRC as well. We show that the CF scheme achieves total exchange rates within a finite-bit gap of the exchange capacity for any number of clusters and users. This limited rate loss is due to noise forwarding from the relay terminal to the users; however, its negative effect becomes less important as the number of users in each cluster increases since the relative strength of the noise variance diminishes.

We also extend the nested lattice coding scheme to the pairwise data exchange model with multiple clusters, and show that employing nested lattice codes yields total exchange rates within a finite-bit gap of the total exchange capacity for any number of clusters. Using structured codes allows the relay to decode only a function of the users' messages rather than decoding each one of them, which fits well with the data exchange model considered here.

The study of multi-way channels for data exchange dates back to Shannon's work on two-way channels [7]. In [8], a multiuser extension of the two-way channel model is studied. These models do not include a relay terminal. The two-way relay channel (TRC), also known as the bidirectional relay channel, models the relay network with two-users exchanging information over a relay terminal. TRC has received considerable attention recently, see [4], [6], [9], [10], [11], [12], [13], [14] and the references therein. The mRC model generalizes the TRC model; in the special case of $L = 1$ and $K = 2$, this model reduces to the TRC.

In [15], multiple simultaneous data transmissions over a relay network is studied with joint network and superposition coding. In an independent work closely related to ours [16], Cui et al. consider in particular the full data exchange model, and study AF, DF and CF schemes. Following our initial study [17], Ong et al. also studied the full date exchange model and characterized the exact capacity region for finite field channels [18]. The pairwise data exchange model is studied from the perspective of optimal power allocation for the special case with orthogonal channels in [19] and bit error rate analysis for interference limited scenarios in [20]. The multi-pair TRC, which corresponds to $L$ clusters with $K = 2$, is also studied in [21].

The following notation and definitions will be used throughout the paper. We denote the set $\{1, \ldots, K\}$ by $\mathcal{I}_K$ for a positive integer $K$. For $l \in \mathcal{I}_K$, we have $\mathcal{I}_K \backslash l = \{1, \ldots, l-1, l+1, \ldots, K\}$. We denote the sequence $(X_1, \ldots, X_n)$ by $X_1^n$. We use $\subset$ for a proposer subset, i.e., $A \backslash B$ is nonempty for any set $B \subset A$, while $\subseteq$ is used for any subset. We define the function $[x]^+ \triangleq \max\{0, x\}$. We also define the function $C(x)$ for a non-negative real number $x$ as

$$C(x) \triangleq \frac{1}{2} \log(1 + x).$$

The rest of the paper is organized as follows. The system model is introduced in Section II. A cut-set outer bound and inner bounds achievable by AF, DF and CF schemes are presented in Section III for the general model. Section IV focuses on the achievable exchange rate for a symmetric Gaussian network. Two special types of networks, the full data exchange and pairwise data exchange models, are studied in Section V and numerical results are provided for these cases. We close the paper with some conclusions highlighting our main results and the insights they provide.

## II. SYSTEM MODEL

We consider a Gaussian mRC in which multiple users exchange messages with the help of a relay terminal, $R$. Users cannot overhear each other's transmissions, hence the relay is essential for communication. We consider full-duplex communication, that is, all terminals including the relay can receive and transmit simultaneously. There are $L \geq 1$ clusters of nodes in the network, where each cluster has $K \geq 2$ users. Users in cluster $j$, $j \in \mathcal{I}_L$, are denoted by $T_{j1}, \ldots, T_{jK}$ (see Fig. 2). $W_{ji} \in \mathcal{W}_{ji}$ is the message of user $T_{ji}$, and user $T_{ji}$ wants to decode the messages $(W_{j1}, \ldots, W_{jK})$ for $j \in \mathcal{I}_L, i \in \mathcal{I}_K$, i.e., the messages of all the users in its own cluster. We denote the set of users in cluster $j$ by $\mathcal{T}_j$, and the set of all users by $\mathcal{T}$.

The Gaussian mRC is modeled as

$$Y_r[t] = \sum_{j=1}^{L}\sum_{i=1}^{K} X_{ji}[t] + Z_r[t] \tag{1}$$

and

$$Y_{ji}[t] = X_r[t] + Z_{ji}[t], \qquad j \in \mathcal{I}_L \text{ and } i \in \mathcal{I}_L \tag{2}$$

where $X_{ji}[t]$ and $Y_{ji}[t]$ are the input and the output at user $T_{ji}$ at time $t$, respectively, while $X_r[t]$ and $Y_r[t]$ are the input and output at the relay, respectively. $Y_r[t]$ is the received signal at the relay and $Y_{ji}[t]$ is the received signal at user $T_{ji}$. $Z_r$ is a zero-mean Gaussian noise at the relay with variance $N_r$, i.e., $Z_r \sim \mathcal{N}(0, N_r)$, and $Z_{ji}$ is the Gaussian noise at user $T_{ji}$, where $Z_{ji} \sim \mathcal{N}(0, N_{ji})$ for $j \in \mathcal{I}_L, i \in \mathcal{I}_K$. All noise variables are independent of each other and the channel inputs and independent and identically distributed (i.i.d.) over time. Average power constraints apply on the transmitted signals at the relay and at the users $T_{ji}$ for all $j \in \mathcal{I}_L$ and $i \in \mathcal{I}_L$:

$$\frac{1}{n}E\left[\sum_{t=1}^{n} |X_r[t]|^2\right] \leq P_r \tag{3}$$

and

$$\frac{1}{n}E\left[\sum_{t=1}^{n} |X_{ji}[t]|^2\right] \leq P_{ji}. \tag{4}$$

Furthermore, although we have a full-duplex operation, the effect of the transmitted signal of each user on its received signal is ignored since it is known at the transmitter, and hence can be subtracted.

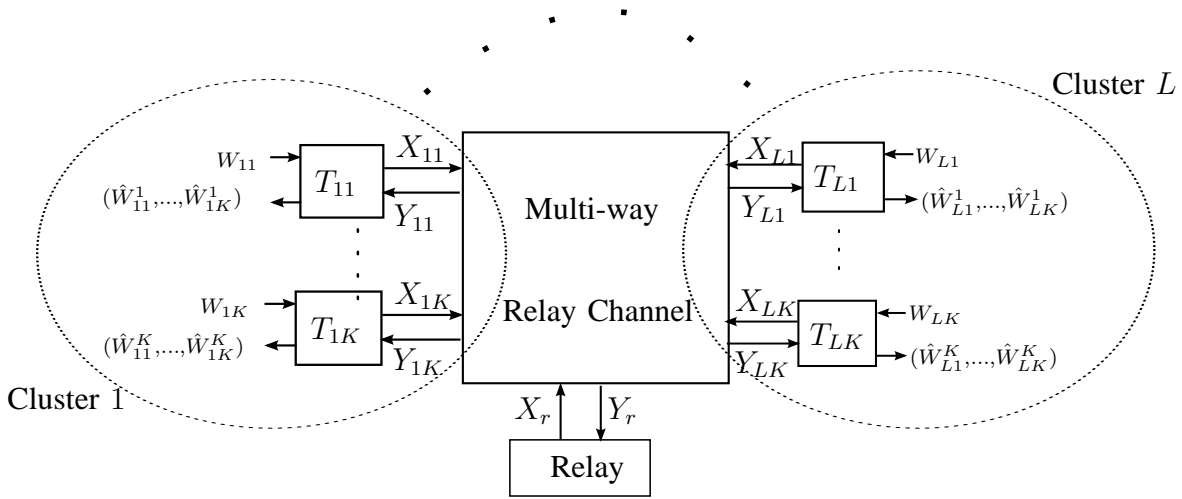

Fig. 2. The mRC model with $L$ clusters, each of which is composed of $K$ distinct terminals. All terminals in a cluster want to receive the messages of all the other terminals in the same cluster. The relay terminal facilitates the data exchange between the terminals.

As in many of the previous work [4], [7], [12], we consider "restricted encoders" at the terminals such that the encoders cannot use their received signals for encoding, and hence, their channel input depends only on their messages. Naturally, the achievable coding schemes proposed in this paper apply to the case without restricted encoders as well; however, our outer bound, and hence, the finite-bit gap arguments are only valid under this restriction.

A $(2^{nR_{11}}, \ldots, 2^{nR_{1K}}, \ldots, 2^{nR_{L1}}, \ldots, 2^{nR_{LK}}, n)$ code for the mRC consists of $N = LK$ sets of integers $\mathcal{W}_{ji} = \{1, 2, \ldots, 2^{nR_{ji}}\}$ for $j \in \mathcal{I}_L$ and $i \in \mathcal{I}_K$ as the message sets, $N$ encoding functions $f_{ji}$ at the users such that

$$x_{ji} = f_{ji}(W_{ji}),$$

a set of encoding functions $\{f_r^t\}_{t=1}^n$ at the relay such that

$$x_{r,t} = f_r^t(Y_r^{t-1}), \quad 1 \leq t \leq n,$$

and $N$ decoding functions at the users:

$$g_{ji} : \mathcal{Y}_{ji}^n \times \mathcal{W}_{ji} \to (\mathcal{W}_{j1}, \ldots, \mathcal{W}_{jK}),$$

such that

$$(\hat{W}_{j1}^i, \ldots, \hat{W}_{jK}^i) = g_{ji}(W_{ji}, Y_{ji}^n).$$

The average probability of error for this system is defined as

$$P_e^n \;=\; \Pr \bigcup_{j\in\mathcal{I}_L, i\in\mathcal{I}_L} \{g_{ji}(W_{ji}, Y_{ji}^n) \neq (W_{j1},\ldots,W_{jK})\}.$$

Observe that the condition $P_e^n \to 0$ implies that individual average error probabilities also go to zero. We assume that the messages $W_{ji}$, $j \in \mathcal{I}_L$, $i \in \mathcal{I}_L$, are chosen independently and uniformly over the message sets $\mathcal{W}_{ji}$. We define the rate vector for cluster $j$ as $\mathbf{R}^j = (R_{j1},\ldots,R_{jK})$ for $j = 1,\ldots,L$ and denote the overall rate vector as $\mathbf{R}^{L,K} = (\mathbf{R}^1,\ldots,\mathbf{R}^L)$.

*Definition 1:* A rate tuple $\mathbf{R}^{L,K}$ is said to be *achievable* for an mRC with $L$ clusters of $K$ users each if there exists a sequence of

$$(2^{nR_{11}},\ldots,2^{nR_{1K}},\ldots,2^{nR_{L1}},\ldots,2^{nR_{LK}}, n)$$

codes such that $P_e^n \to 0$ as $n \to \infty$. The corresponding *capacity region* is the convex closure of all achievable rate tuples.

## III. THE ACHIEVABLE RATE REGION

In this section we provide inner and outer bounds on the capacity region of the Gaussian mRC. The outer bound is based on the combination of the classical cut-set bound [22] and a genie-aided bound. The proposed achievable rate regions are based on the relaying schemes originally developed for the classical one-way relay channel. In particular, we consider AF, DF and CF schemes, and identify the corresponding achievable rates. Unlike the classical one-way relay channel, in the multi-way relay setting, the transmitters can exploit the knowledge of their own transmit signals to improve rate region.

### A. Outer bound

To derive an outer bound on the capacity region of the mRC, we first consider the cut-set bound. Choose any proper subset of users from each cluster, and let $\mathcal{S}_j \subset \mathcal{I}_K$ denote the set of users chosen from cluster $j$. We will consider the information flow from these users to the remaining users. Consider the cut formed by $\bigcup_{j=1}^{L} \mathcal{S}_j$, which forms a MAC to the relay and provides the following cut-set bound on the total rate that needs to be transmitted over this cut:

$$\sum_{j=1}^{L} \sum_{i\in\mathcal{S}_j} R_{ji} \leq C\left(\frac{\sum_{j=1}^{L}\sum_{i\in\mathcal{S}_j} P_{ji}}{N_r}\right) \tag{5}$$

9for all $\mathcal{S}_j \subset \mathcal{I}_K$.

Next we consider a genie-aided outer bound. Choose one user from each cluster. Assume that a genie provides to the remaining users and the relay all the messages in the network. Hence, the network only needs to transmit the messages in their clusters to the set of chosen users. Since relay already knows all the messages, the setup boils down to a broadcast channel from the relay to the set of chosen users. Note here that, while the previous cut-set outer bound is valid for non-restricted encoders as well, this genie-aided outer bound is based on the restricted encoder assumption, and hence, ignores the potential feedback signal that can be transmitted from the receivers.

The capacity region of a Gaussian broadcast channel with $L$ receivers, power constraint $P_r$ at the transmitter, and noise variances $N_1, \ldots, N_L$ at the receivers is given as [22]

$$\mathcal{C}^{BC}(P_r, N_1, \ldots, N_L) \triangleq \left\{ (R_1, \ldots, R_L) : 0 \leq R_j \leq C\left(\frac{\alpha_j P_r}{P_r \sum_{k=1}^{L} \alpha_k \mathbb{1}_{[N_k < N_j]} + N_j}\right), \right.$$
$$\left. \alpha_j \geq 0 \text{ for } j \in \mathcal{I}_L, \sum_{j=1}^{L} \alpha_j = 1 \right\}, \quad (6)$$

where $\mathbb{1}_x = 1$ if $x$ is true and $0$ otherwise. Assuming that user $T_{jl_j}$, $l_j \in \mathcal{I}_K$, $j \in \mathcal{I}_L$ is chosen in cluster $j$, the total rate of the messages to this receiver is $R_{l_j}^j \triangleq \sum_{i \in \mathcal{S}_j} R_{ji}$ where $W_j \triangleq \mathcal{I}_K \setminus \{l_j\}$. These rates need to be supported by the broadcast channel, i.e. we need:

$$(R_{l_1}^1, \ldots, R_{l_L}^L) \in \mathcal{C}^{BC}(P_r, N_{l_1}, \ldots, N_{l_j}) \quad (7)$$

for all choices of $l_1, \ldots, l_L$, where $l_j \in \mathcal{I}_K$ and $j \in \mathcal{I}_L$. The intersection of the bounds in (5) and (7) provides us an outer bound on the capacity region of the mRC.

Note that this is a capacity region outer bound for the case of restricted encoders as we ignored the feedback to the encoders.

*B. Amplify-and-forward (AF) Relaying*

In AF relaying, the relay terminal amplifies its received signal within its power constraint and broadcasts to the receivers. However, since the signals from users that belong to different clusters act as noise to each other, we consider time-sharing among clusters, and apply the AF strategy separately for each cluster within its own timeslot. Let $\tau_j$ denote the portion of the channel allocated to cluster $j$ with $\sum_{j=1}^{L} \tau_j = 1$. Within the timeslot of each cluster, all the

DRAFT

users in that cluster transmit, and the relay scales its received signal and broadcasts to the users. Within the timeslot for cluster $j$ the relay's transmit signal is given by

$$X_r^j = \sqrt{\frac{P_r^j}{\sum_{i=1}^{K} P'_{ji} + N_r}}(X_{j1} + \cdots + X_{jK} + Z_r),$$

where $P'_{ji}$ is the transmit power of user $T_{ji}$ and $P_r^j$ is the transmit power of the relay at timeslot $j$. We have $0 \leq P'_{ji} \leq \frac{P_{ji}}{\tau_j}$ and $\sum_{j=1}^{L} \tau_j P_r^j \leq P_r$. Each user can cancel out the effect of its own transmit signal, and decodes the messages of the other users in its own cluster. Since the transmission from a user acts as noise on the other users' transmissions, users do not necessarily transmit at full power. At each receiver, we have a Gaussian MAC with $K-1$ users, and we assume Gaussian codebooks are used.

*Proposition 1:* For a Gaussian mRC with $L$ clusters of $K$ users each, the rate region characterized by the union of the rate tuples satisfying the following inequalities is achievable with AF relaying and time-sharing between clusters

$$0 \leq \sum_{k \in \mathcal{S}} R_{jk}^{AF} \leq \tau_j C\left(\frac{\sum_{k \in \mathcal{S}} P'_{jk}}{N_r + \frac{\sum_{i=1}^{K} P'_{ji} + N_r}{P_r^j} N_{jl}}\right), \quad (8)$$

for all $j \in \mathcal{I}_L$, $l \in \mathcal{I}_K$ and $\mathcal{S} \subseteq \mathcal{I}_K \backslash l$ such that $0 \leq P'_{ji} \leq \frac{P_{ji}}{\tau_j}$ for all $j \in \mathcal{I}_L$, $i \in \mathcal{I}_K$, $\sum_{j=1}^{L} \tau_j P_r^j \leq P_r$, $\tau_j \geq 0$ for $j = 1, \ldots, L$ and $\sum_{j=1}^{L} \tau_j = 1$.

*C. Decode-and-forward (DF) Relaying*

In DF relaying, the relay decodes messages from all the users, and broadcasts each message to all its recipients. DF consists of two transmission phases: the first phase is the MAC from the users to the relay, and the second phase is the broadcast channel from the relay to the users. Note that, due to the full-duplex nature of the relay operation, these two phases occur simultaneously for consecutive message blocks. The messages of all users can be decoded at the relay at the end of the multiple access phase if

$$\sum_{j \in \mathcal{S}_1} \sum_{i \in \mathcal{S}_2} R_{ji}^{DF} \leq C\left(\frac{\sum_{j \in \mathcal{S}_1} \sum_{i \in \mathcal{S}_2} P_{ji}}{N_r}\right), \quad (9)$$

for all $W_1 \subseteq \mathcal{I}_L$, $W_2 \subseteq \mathcal{I}_K$.

In the broadcast phase, we consider time-sharing among clusters, that is, the relay divides the channel block into $L$ timeslots proportional to $\tau_j \geq 0$ where $\sum_{j=1}^{L} \tau_j = 1$. For $j \in \mathcal{I}_L$, the relay broadcasts the messages $W_{j1}, \ldots, W_{jK}$ to users $T_{j1}, \ldots, T_{jK}$ within the $j$-th timeslot. For broadcasting within the $j$-th timeslot, rather than broadcasting each message one by one to its intended receivers, the relay broadcasts all the messages simultaneously to all the receivers by using the coding scheme introduced in [2], which exploits the availability of the users' own messages in decoding the remaining messages.

In [2], Tuncel has considered broadcasting a source to multiple receivers each of which has its own correlated side information, and characterized necessary and sufficient conditions for the reliable transmission of the source to all the receivers. In our setting, we consider $W_{j1}, \ldots, W_{jK}$ as the source message within timeslot $j$ and $W_{ji}$ as the correlated side information at user $T_{ji}$ for $j \in \mathcal{I}_L$, $i \in \mathcal{I}_K$. In this coding scheme, the relay generates a codebook of size $2^{nR_{j1}} \times 2^{nR_{j2}} \times \cdots \times 2^{nR_{jK}}$ for each cluster $j$, consisting of $\tau_j n$-length codewords i.i.d. Gaussian $\mathcal{N}(0, P_r^j)$, where $\sum_{j=1}^{L} \tau_j P_r^j \leq P_r$ and $\sum_{j=1}^{L} \tau_j = 1$. For each message combination $(W_{j1}, \ldots, W_{jK})$ the relay transmits the corresponding codeword over the channel. Each receiver finds the message indices by joint typicality using its channel output and its own message, which acts as the side information in our model. The analysis of the coding scheme follows from [2]. This coding scheme is also used in [23], [24] and [25] for identifying the capacity region of broadcast channels with message side information. We can show that the messages can be decoded by all the users if

$$\sum_{i \in \mathcal{I}_K \setminus \{l\}} R_{ji}^{DF} \leq \tau_j C \left( \frac{P_r^j}{N_{jl}} \right), \tag{10}$$

for all $j \in \mathcal{I}_L$ and $l \in \mathcal{I}_K$.

*Proposition 2:* For a Gaussian mRC with $L$ clusters of $K$ users each, the rate region characterized by the union of the rate tuples satisfying the following inequalities is achievable with DF relaying:

$$\sum_{j \in \mathcal{S}_1} \sum_{i \in \mathcal{S}_2} R_{ji}^{DF} \leq C \left( \frac{\sum_{j \in \mathcal{S}_1} \sum_{i \in \mathcal{S}_2} P_{ji}}{N_r} \right), \text{ for all } W_1 \subseteq \mathcal{I}_L, W_2 \subseteq \mathcal{I}_K, \tag{11}$$

$$\tag{12}$$

and

$$\sum_{i \in \mathcal{I}_K \setminus \{l\}} R_{ji}^{DF} \leq \tau_j C \left( \frac{P_r^j}{N_{jl}} \right), \text{ for all } j \in \mathcal{I}_L, l \in \mathcal{I}_K, \tag{13}$$

such that $\sum_{j=1}^{L} \tau_j P_r^j \leq P_r$, $\tau_j \geq 0$ for $j = 1, \ldots, L$ and $\sum_{j=1}^{L} \tau_j = 1$.

## D. Compress-and-forward (CF) Relaying

Next, we consider CF relaying which was introduced in [1] for the 'one-way' single relay channel. In the CF scheme in a one-way relay channel, the relay transmits a quantized version of its received signal to the destination. Since the destination has its own received version of the source signal, which is correlated with the relay's received signal, the relay exploits this correlated side information at the receiver by using Wyner-Ziv compression [26]. Then the destination combines its received signal and the quantized version of the relay's received signal to decode the underlying source message.

Note that in our mRC setup the users do not overhear each other's signals; however, they still have access to the side information correlated with the relay's signal: their own transmit signals. Therefore, we propose a transmission scheme for the multi-way relay channel based on CF relaying that exploits this side information at the users.

In the CF scheme proposed in [4] for the TRC and extended in [17] to the mRC, the relay terminal quantizes its received signal and broadcasts this quantized channel output to the users. Hence, similarly to the DF scheme, the scenario is equivalent to broadcasting a common source to multiple receivers with correlated side information. However, note that, unlike the DF case, here we are interested in broadcasting a quantized version of the relay's received signal, rather than its lossless transmission. Yet it is possible to employ a coding scheme similar to the one used for DF relaying to exploit the side information at the users simultaneously. Improvement in the achievable rates is possible by employing layered digital codes as in [3], or by further exploiting analog transmission as in [27]. However, this type of CF scheme, despite not using explicit binning, still requires decoding of the quantized relay signal at the users, which is not a requirement of the problem. Instead, the users can directly decode the messages of the other users without decoding the quantized relay signal. This coding scheme is originally considered in [12] for the TRC, and recently generalized to multiple relay networks in [28]. A variation of the CF scheme in which the receivers decode only the bin indices rather than the compression indices before decoding the message indices is studied in [29].

We first provide an achievable rate region for a general discrete memoryless channel in which the channel from the users to the relay is characterized by the conditional probability distribution $p(y_r|x_1, \ldots, x_K)$ and the channel from the relay to the users is characterized by $p(y_1, \ldots, y_K|x_r)$.

Note in this model that, the channel output at a terminal does not depend on the channel input of that terminal. This is in accordance with the Gaussian model, in which case the known channel input can be subtracted from the output of each user. We consider a single cluster to simplify the rate region expression and drop the cluster index in the random variables. Later, we use this expression to obtain an achievable rate region for the Gaussian model with multiple clusters.

*Theorem 1:* For a discrete memoryless mRC with $K$ users exchanging information among each other, the rate tuples satisfying the following inequalities are achievable by CF if,

$$\sum_{k \in \mathcal{S}} R_k^{CF} \leq \min \left\{ I(X(\mathcal{S}); \hat{Y}_r | X(\mathcal{S}^c), Q), \left[ \min_{t \in \mathcal{S}^c} I(X_r; Y_t | Q) - I(Y_r; \hat{Y}_r | X^K, Q) \right]^+ \right\}, \quad (14)$$

for all $\mathcal{S} \subset \mathcal{I}_K$, for some probability distribution in the form

$$p(q)p(x_1|q) \cdots p(x_K|q)p(x_r|q)p(y_r|x_1, \ldots, x_K)p(\hat{y}_r|y_r)p(y_1, \ldots, y_K).$$

*Proof:* See Appendix A for the details. ∎

*Remark 1:* In our achievable coding scheme, the users transmit a new message in every channel block and the relay quantizes and forwards its observation without Wyner-Ziv binning as in [3]. Each destination decodes the messages of the remaining users by joint typicality after receiving the signal transmitted by the relay at each channel block. The users directly decode the message indices without trying to decode the quantized relay codeword first. While repetition coding and joint decoding is considered in [28], our result illustrates that this is not needed in the mRC network setup considered here. A similar result is recently obtained for a single source-single destination multiple relay network in [30].

In the Gaussian setup, as in AF, we consider time-sharing among the user clusters in the multiple access phase as well as in the broadcast phase. This will prevent multiple user clusters from interfering with each other's signals, which would decrease the quality of the quantized signal broadcasted by the relay.

*Proposition 3:* For a Gaussian mRC with $L$ clusters of $K$ users each, the rate tuples satisfying the following inequalities are achievable by CF:

$$\sum_{k \in \mathcal{S}_j} R_{jk}^{CF} < \tau_j \min \left\{ C\left( \frac{\sum_{k \in \mathcal{S}_j} P'_{jk}}{N_r + N_Q^j} \right), C\left( \frac{P_r^j}{\max_{t \in \mathcal{S}^c} N_{jt}} \right) - C\left( \frac{N_r}{N_Q^j} \right) \right\}. \quad (15)$$

for all $j \in \mathcal{I}_L$ and $S_j \subset \mathcal{I}_K$ and some $N_Q^j > 0$, such that $P'_{ji} \leq \frac{P_{ji}}{\tau_j}$ for all $j \in \mathcal{I}_L$ and $i \in \mathcal{I}_K$, $\sum_{j=1}^{L} \tau_j P_r^j \leq P_r$, $\tau_j \geq 0$ for $j = 1, \ldots, L$ and $\sum_{j=1}^{L} \tau_j = 1$.

DRAFT

*Proof:* We apply time-sharing between clusters, hence the achievable CF rates for each cluster is scaled with the portion of time $\tau_j$ allocated to that cluster. We also allow the relay to allocate its power among various clusters. We let the users generate Gaussian codebooks. In particular, we let $X_{ji} \sim \mathcal{N}(0, P'_{ji})$ and $X_r \sim \mathcal{N}(0, P_r)$, where $P'_{ji} \leq P_{ji}$. Without claiming optimality, we also let the quantization noise at the relay for quantizing the received signal $Y_r^j$ for cluster $j$ be Gaussian, that is,

$$\hat{Y}_r^j = Y_r^j + Q_j, \tag{16}$$

where $Q_j$ is a Gaussian random variable with $Q_j \sim \mathcal{N}(0, N_Q^j)$, $N_Q^j > 0$, and independent of $Y_r^j$. Calculating the mutual information expressions for these Gaussian random variables results in the above rate region. ∎

## E. Lattice Coding

In the previous sections, we have concentrated on various random coding schemes for communication over the mRC. Recently, it has been shown in [11], [6], [31] and [32], that nested lattice codes can be effective in achieving higher rates in some Gaussian networks by exploiting the topology of the network. Basic motivation in employing lattice codes in these architectures is to allow the relay nodes to decode only the modulo sum of the messages rather than decoding the individual messages.

Unfortunately this structured coding scheme does not directly scale with increasing number of simultaneously transmitting users at each instant, that is, by knowing the modulo sum of more than two messages and only one of the messages, the users cannot decode the remaining messages. Hence, in our setup, we concentrate on the use of lattice codes for the case with $K = 2$. This model is equivalent to having multiple two-way relay channels served simultaneously by a single relay terminal [20], [19]. We term this model the *mRC with pairwise data exchange*.

In this section, we provide an achievability scheme based on nested lattice codes [33], in which each user in the same pair uses a lattice structure to transmit its messages so that the addition of any two message points is also a member of the lattice. The relay terminal decodes the modulo sum of the transmitted lattice points, and then broadcasts this modulo sum to both users, each of which can decode the other user's message by subtracting its own message.

We next provide a brief review of nested lattice coding that will be required for the presentation of the coding scheme (see [33] or [34] for further details). An $n$-dimensional lattice $\Lambda$ is defined as

$$\Lambda = \{GX : X \in \mathbb{Z}^n\},$$

where $G \in \mathbf{R}^n$ is the generator matrix. For any $x \in \mathbb{R}^n$, the quantization of $X$ maps $X$ to the nearest lattice point in Euclidean distance:

$$Q_\Lambda(X) \triangleq \arg\min_{Q \in \Lambda} \|X - Q\|.$$

The mod operation is defined as

$$X \mod \Lambda = X - Q_\Lambda(X).$$

The fundamental Voronoi region $\mathcal{V}(\Lambda)$ is defined as $\mathcal{V}(\Lambda) = \{X : Q_\Lambda(X) = 0\}$, whose volume is denoted by $V(\Lambda)$ and is defined as $V(\Lambda) = \int_{\mathcal{V}(\Lambda)} dX$. The second moment of a lattice $\Lambda$ is given by

$$\sigma^2(\Lambda) = \frac{1}{nV(\Lambda)} \int_{\mathcal{V}(\Lambda)} \|X\|^2 dX,$$

while the normalized second moment is defined as

$$G(\Lambda) = \frac{\sigma^2(\Lambda)}{V(\Lambda)^{2/n}}.$$

We use a nested lattice structure as in [5], where $\Lambda_c$ denotes the coarse lattice and $\Lambda_f$ denotes the fine lattice and we have $\Lambda_c \subseteq \Lambda_f$. All transmitters use the same coarse and fine lattices for coding. We consider lattices such that $G(\Lambda_c) \approx \frac{1}{2\pi e}$ and $G(\Lambda_f) \approx \frac{1}{2\pi e}$, whose existence is shown in [5]. In nested lattice coding, the codewords are the lattice points of the fine lattice that are in the fundamental Voronoi region of the coarse lattice. Moreover, we choose the coarse lattice (i.e., the shaping lattice) such that $\sigma^2(\Lambda_c) = P$ to satisfy the power constraint. The fine lattice is chosen to be good for channel coding, i.e., it achieves the Poltyrev exponent [5].

We assume that both of the users $T_{j1}$ and $T_{j2}$ in pair $j$ use the same nested lattice structure for coding, and hence, achieve the same rate $R_j^{lattice}$. We also assume that both users have the same power constraint $P_j$, as additional power at one of the users would be useless in the proposed scheme. However, we want to note here that it is also possible to combine this lattice coding scheme with a random coding scheme as in [35] such that the user with more power available can superimpose an additional random code on top of the lattice code, and hence, achieve a higher data rate.

We use time division among the user pairs in the transmission to the relay terminal. Each transmitter $T_{ji}$ maps its message $W_{ji}$ to a fine lattice point $V_{ji} \in \Lambda_f \cap \mathcal{V}(\Lambda_c)$, $j \in \mathcal{I}_L$ and $i = 1, 2$. Each user employs a dither vector $U_{ji}$ which is independent of the dither vectors of the other users and of the messages and is uniformly distributed over $\mathcal{V}(\Lambda_c)$. We assume all the terminals in the network know the dither vectors. The transmitted codeword from transmitter $T_{ji}$ is given by

$$X_{ji} = (V_{ji} - U_{ji}) \mod \Lambda_c.$$

It can be shown that $X_{ji}$ is also uniform over $\mathcal{V}(\Lambda_c)$.

The relay decodes the modulo sums of the messages, $V_j \triangleq (V_{j1} + V_{j2}) \mod \Lambda_c$, instead of decoding individual messages. Due to the group structure of the lattice, $V_j$ also belongs to the fine lattice. Moreover, it is possible to show that $V_j$ is also uniformly distributed over the fine lattice points within the Voronoi region of the coarse lattice, i.e., over the set $\Lambda_f \cap \mathcal{V}(\Lambda_c)$.

Following [33] and [11], it is possible to show that there exist nested lattices at rates arbitrarily close to

$$R_j^{lattice} = \tau_j C^+ \left( \frac{P_j}{\tau_j N_r} - \frac{1}{2} \right), \tag{17}$$

where $\sum_{j=1}^L \tau_j = 1$ and $C^+(x) = C(x)$ if $x \geq 1$ and $0$ otherwise. This allows the relay to decode $V_j$'s with vanishing error probability.

For the broadcasting of the modulo sums from the relay to the pairs, the rate is bounded by the rate that can be transmitted to each user, i.e., we need $(R_1^{lattice}, \ldots, R_L^{lattice}) \in \mathcal{C}^{BC}(P_r, N_1, \ldots, N_L)$, where $N_j \triangleq \max\{N_{j1}, N_{j2}\}$.

## IV. EXCHANGE RATE FOR A SYMMETRIC NETWORK

In this section, we focus on a symmetric network with equal power constraints at the users, $P_{ji} = P$, and compare the achievable equal rate points with the proposed relaying schemes, i.e., $R_{ji} = R$ for all $j \in \mathcal{I}_L$ and $i \in \mathcal{I}_L$. Exchange rate analysis will allow us to compare these schemes analytically for different numbers of clusters and users and with different power constraints. We say that a total exchange rate of $R_t$ is achievable for a system with $L$ clusters 4and $K$ users in each cluster if $(\frac{R_t}{LK}, \ldots, \frac{R_t}{LK})$ is an achievable rate tuple. The exchange capacity is defined as the supremum of all achievable total exchange rates, i.e.,

$$C_{sym}^{L,K} \triangleq \sup\{LKR : (R, \ldots, R) \text{ is achievable}\}.$$

We find lower and upper bounds on the exchange capacity of the network. In general, these bounds do not match and the exchange capacity of the mRC remains open. However, we show below that the gap between these two is less than a finite number of bits which is independent of the power constraints of the users. In the analysis of the exchange capacity, to simplify the notation and to focus on the fundamental behavior of the analyzed schemes, we consider a symmetric network with $P_{ji} = P$ and $N_r = N_{ji} = 1$ for all $j \in \mathcal{I}_L$, $i \in \mathcal{I}_L$.

We start with the upper bound on the exchange capacity. For a symmetric Gaussian mRC with $L$ clusters of $K$ users each, the exchange capacity is upper bounded by

$$R_{UB}^{L,K} = \frac{K}{K-1} \min\{C(L(K-1)P), C(P_r)\}. \tag{18}$$

With AF relaying, the achievable total exchange rate is found as follows from Proposition 1 by letting $\tau_j = 1/L$, $P_r^j = P_r$ and $P'_{ji} = LP$:

$$R_{AF}^{L,K} = \frac{K}{(K-1)} C\left(\frac{L(K-1)PP_r}{1 + LKP + P_r}\right). \tag{19}$$

In a symmetric Gaussian mRC with $L$ clusters of $K$ users each, the following exchange rate is achievable with DF relaying by letting $\tau_j = 1/L$ and $P_r^j = P_r$:

$$R_{DF}^{L,K} = \min\left\{C(LKP), \frac{K}{K-1}C(P_r)\right\}. \tag{20}$$

*Remark 2:* Comparing (20) and (18), we can show that DF achieves the exchange capacity when

$$P_r \leq (1 + LKP)^{1-\frac{1}{K}} - 1.$$

This corresponds to the case in which the relay power is the bottleneck, i.e., the exchange capacity is limited by the rate at which the relay can broadcast to the users. The range of $P_r$ for which DF is optimal increases as the number of clusters, the number of users within each cluster, or the power constraint $P$ of the users increases.

Finally, the total exchange rate achievable by CF over a symmetric network is given by

$$R_{CF}^{L,K} = \frac{K}{K-1} C\left(\frac{L(K-1)PP_r}{1 + L(K-1)P + P_r}\right). \tag{21}$$

17DRAFT

*Remark 3:* Comparing (19) and (21), we observe that, for an arbitrary number of clusters and terminals within each cluster ($L \geq 1, K \geq 2$), the total exchange rate achieved by AF is lower than CF. Yet, we remark that the simplicity of AF relaying compared to CF may be attractive in practice. Moreover, the gap between the two is upper bounded:

$$R_{CF}^{L,K} - R_{AF}^{L,K} \leq \frac{K}{2(K-1)} \log(\frac{K}{K-1}), \tag{22}$$

which is independent of the power constraints and the number of clusters.

In the next theorem, we prove that the CF protocol achieves rates within a constant number of bits of the exchange capacity for an arbitrary number of clusters and users independent of the available power at the users and the relay.

*Theorem 2:* For a symmetric Gaussian mRC with $L$ clusters of $K$ users each, the CF protocol achieves rates within $\frac{K}{2(K-1)}$ bits of the exchange capacity.

*Proof:* First, assume that $P_r \geq L(K-1)P$. Then we have the following chain of inequalities:

$$R_{CF}^{L,K} = \frac{K}{K-1} C\left(\frac{L(K-1)PP_r}{1+L(K-1)P+P_r}\right) \tag{23}$$

$$= \frac{K}{2(K-1)} \log\left(1 + \frac{L(K-1)PP_r}{1+L(K-1)P+P_r}\right) \tag{24}$$

$$= \frac{K}{2(K-1)} \left[\log(1+L(K-1)P) + \log\left(\frac{1+P_r}{1+L(K-1)P+P_r}\right)\right] \tag{25}$$

$$\geq R_{UB}^{L,K} + \frac{K}{2(K-1)} \log\left(\frac{1+P_r}{1+2P_r}\right) \tag{26}$$

$$\geq R_{UB}^{L,K} - \frac{K}{2(K-1)}, \tag{27}$$

where (26) follows from the assumption that $P_r \geq L(K-1)P$.

Next, assuming $P_r \leq L(K-1)P$, we have

$$R_{CF}^{L,K} = \frac{K}{2(K-1)} \left[\log(1+P_r) + \log\left(\frac{1+L(K-1)P}{1+L(K-1)P+P_r}\right)\right] \tag{28}$$

$$\geq R_{UB}^{L,K} + \frac{K}{2(K-1)} \log\left(\frac{1+L(K-1)P}{1+2L(K-1)P}\right)$$

$$\geq R_{UB}^{L,K} - \frac{K}{2(K-1)}. \tag{29}$$

∎

*Remark 4:* It is noteworthy that the constant gap to the capacity is only a function of $K$, and is independent of the number of clusters and the power constraints of the users and the relay. We can conclude that CF is nearly optimal in the high power regime for which the finite bit gap to the capacity becomes negligible. Note that this finite bit gap is bounded by one bit independently of the number of users within each cluster $K$ and decays to half a bit as $K$ increases.

*Remark 5:* A direct consequence of Remark 3 and Theorem 2 is that the AF protocol achieves total exchange rates within $\frac{K(1+\log K)}{2(K-1)}$ bits of the exchange capacity. We can further bound this gap from above by $1 + \log K$, which scales with an increasing number of users within each cluster.

## V. SPECIAL NETWORKS

### A. The Multi-way Relay Channel with Full Data Exchange

In this section we consider a special mRC with a single cluster $L = 1$, that is, each user wants to decode all the messages in the system. We term this model the *mRC with full data exchange*. A similar model, the *multiway channel*, in which there is no relay terminal, and the users can receive each other's signals is considered in [8].

Let us assume that the relay's power scales with the number of users, i.e., $P_r = KP$. In this case we have $R_{UB}^{1,K} = \frac{K \cdot C((K-1)P)}{K-1}$ and $R_{DF}^{1,K} = C(KP)$. We can see from these expressions that, with increasing power, the gap between the two increases and can be arbitrarily large when $P$ is very high. In Fig. 3, we plot the upper bound and the achievable exchange rates for this setup. We see from the plot that the gap between the upper bound and the achievable exchange rate with DF diverges quickly with increasing power especially in the case of small numbers of users. The total exchange rate decreases with the increasing number of users in the system. We have a finite gap between the achievable rate of the CF scheme and the upper bound at all power values. A similar finite bit gap is also observed between the CF and AF schemes as was shown analytically. Especially for a small number of users, the rate of CF dominates the rate of DF for a wide range of power values. On the other hand, DF achieves higher exchange rates than CF in the low power regime. The range of power values in which the DF dominates CF gets larger with the number of users in the system. This is due to the fact that CF forwards more noise when there is increased interference. Similar observations can also be made when the relay power does not scale with the number of users, i.e., $P_r = P$, which is illustrated in Fig. 4.

$X_{LK}$
$Y_{11}$
$Y_{1K}$
$Y_{L1}$
$Y_{LK}$
$W_{11}$
$W_{1K}$
$W_{L1}$
$W_{LK}$
$(\hat{W}_{11}^1,...,\hat{W}_{1K}^1)$
$(\hat{W}_{11}^K,...,\hat{W}_{1K}^K)$
$(\hat{W}_{L1}^1,...,\hat{W}_{LK}^1)$
$(\hat{W}_{L1}^K,...,\hat{W}_{LK}^K)$
Multi-way
Relay Channel
Cluster 1
Cluster $L$

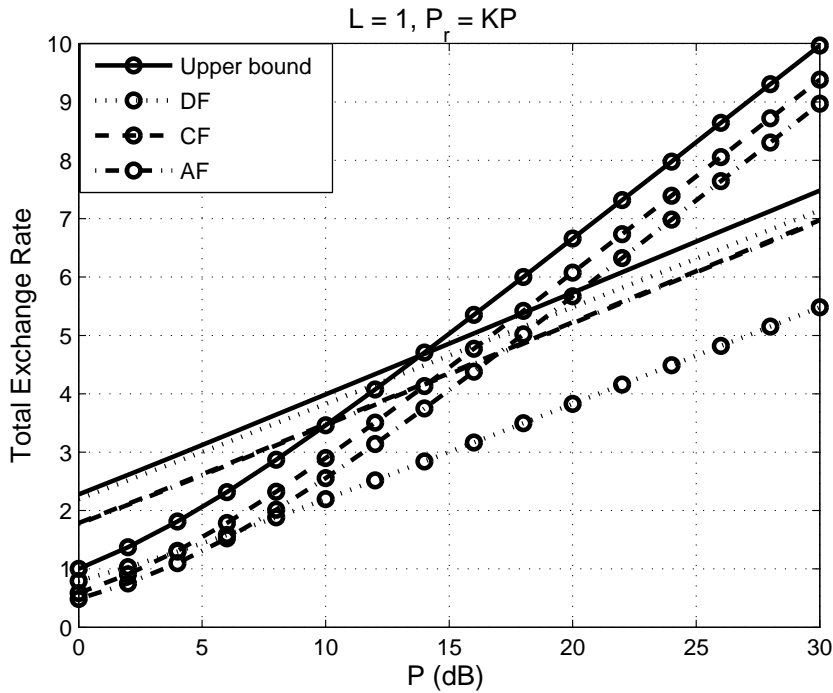

Fig. 3. Total exchange rate versus the user power, $P$. The relay power is equal to the total user power, i.e., $P_r = KP$. We illustrate rates for $K = 2$ (the lines with the marker), and $K = 20$ users.

In Fig. 5, we plot the upper bound and the achievable total exchange rate versus the number of users for the mRC with full data exchange. The lines marked with a circle correspond to the case where the relay power scales with the number of users as $P_r = KP$, while the unmarked lines correspond to the case where the relay power is fixed as $P_r = P$. From Theorem 2, the gap between the upper bound and the achievable total exchange rate with CF for $L = 1$ is $\frac{K}{2(K-1)}$. This gap approaches $0.5$ bits as the number of users $K$ increases independently of the power constraints. We can see that the gap is much smaller when the relay power is equal to the power constraint of each user. With the number of users increasing, both DF and CF get very close to the upper bound. The DF scheme achieves the upper bound with a smaller number of users in the system for the power constraint considered in this figure. In both cases, the achievable rate of the AF scheme is very close to the one achieved by the CF scheme. The gap between the two decreases with the number of users in the system. Note also the initial behavior of the total exchange rate with increasing number of users for the case $P_r = KP$. The rate falls sharply due

$X_{LK}$
$Y_{11}$
$Y_{1K}$
$Y_{L1}$
$Y_{LK}$
$W_{11}$
$W_{1K}$
$W_{L1}$
$W_{LK}$
$(\hat{W}_{11}^1,...,\hat{W}_{1K}^1)$
$(\hat{W}_{11}^K,...,\hat{W}_{1K}^K)$
$(\hat{W}_{L1}^1,...,\hat{W}_{LK}^1)$
$(\hat{W}_{L1}^K,...,\hat{W}_{LK}^K)$
Multi-way Relay Channel
Cluster 1
Cluster $L$

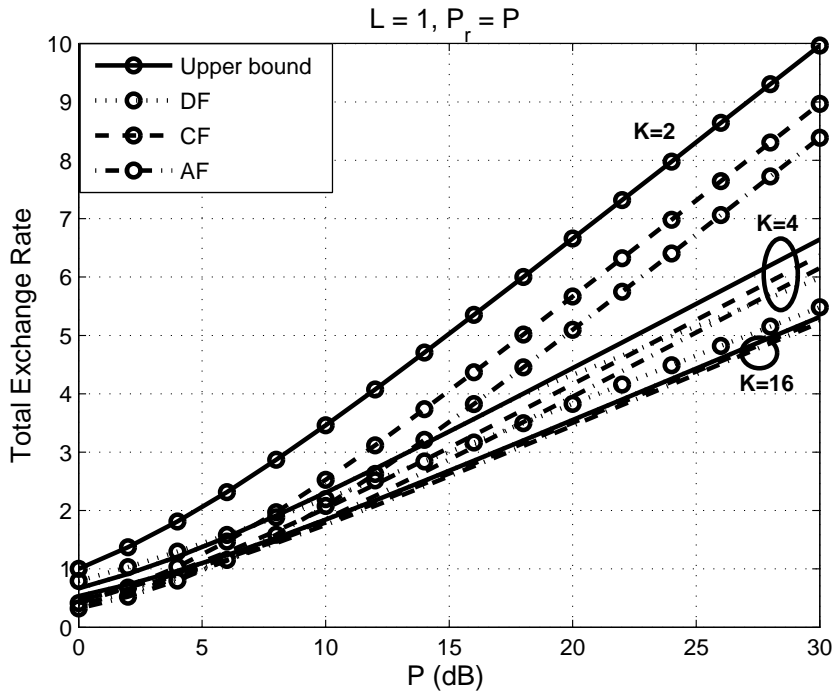

Fig. 4. Total exchange rate versus the user power with $P_r = P$. We illustrate rates for $K = 2$ (the lines with the marker), $K = 4$ and $K = 8$ users.

to the interference introduced by the new users. However, the effect of the interference saturates after a certain number of users, and the total exchange rate starts increasing again. We can prove analytically that the exchange capacity goes to infinity as the number of users goes to infinity if the relay power is scaled with the number of users, whereas it saturates when the relay power is kept constant.

## B. The multi-way Relay Channel with Pairwise Data Exchange

In the previous subsection we focused on full-data exchange, in which case each user wants to decode the messages of all other users. This constitutes one extreme in the mRC model. Another extreme would be to assume that users are paired, and each user is interested only in the data of its partner, i.e., $L \geq 1$ and $K = 2$. For the pairwise data exchange model in addition to the random coding schemes, we also have a nested lattice coding scheme provided in Section III-E.

For the lattice coding scheme in a symmetric network with $L > 1$ clusters, we use time-sharing among the clusters for both the lattice coded multiple access and the broadcast phases. Each

$X_{LK}$
$Y_{11}$
$Y_{1K}$
$Y_{L1}$
$Y_{LK}$
$W_{11}$
$W_{1K}$
$W_{L1}$
$W_{LK}$
$(\hat{W}_{11}^1,...,\hat{W}_{1K}^1)$
$(\hat{W}_{11}^K,...,\hat{W}_{1K}^K)$
$(\hat{W}_{L1}^1,...,\hat{W}_{LK}^1)$
$(\hat{W}_{L1}^K,...,\hat{W}_{LK}^K)$
Multi-way
Relay Channel
Cluster 1
Cluster $L$

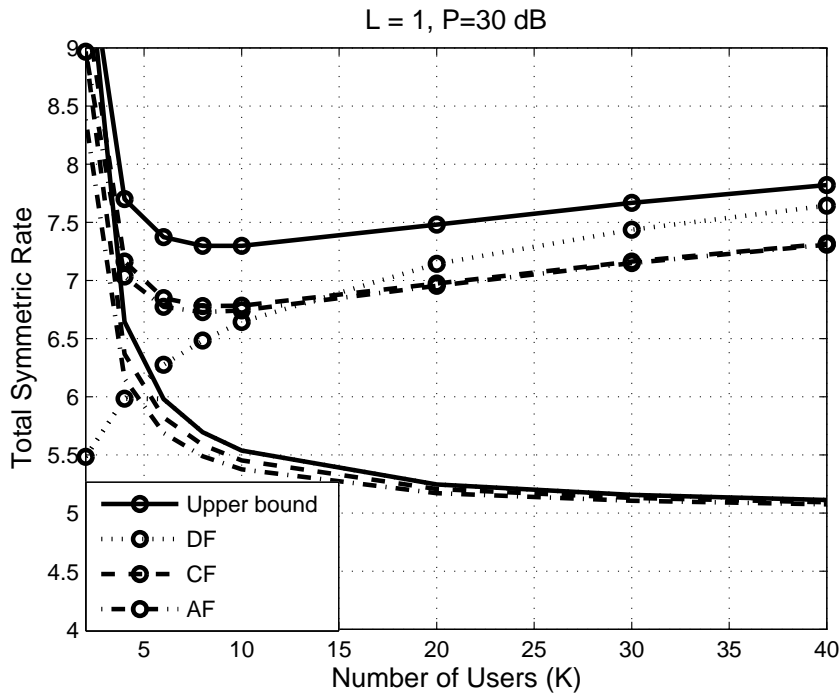

Fig. 5. Achievable total exchange rate versus the number of users with $P = 30$ dB. The lines with the marker correspond to the case with $P_r = KP$, while the nonmarked curves correspond to the case with $P_r = P$.

pair will transmit $1/L$ portion of the timeslot using the same nested lattice code while scaling their power level accordingly. Then the relay broadcasts each pair's modulo sum to both users over $1/L$ portion of the timeslot.

For the broadcasting of the modulo sums from the relay to the pairs, the rate is bounded by $\frac{1}{L}C(P_r)$. Hence, the following exchange rate can be achieved by nested lattice codes:

$$R_{lattice}^{L,2} = \min\left\{\frac{\max\left\{0, C\left(LP - \frac{1}{2}\right)\right\}}{L}, \frac{C(P_r)}{L}\right\}. \qquad (30)$$

*Remark 6:* We can see from (30) that lattice coding achieves the exchange capacity if $0 \leq P_r \leq LP - \frac{1}{2}$. In general, assuming that $LP \geq 1/2$, the total exchange rate achievable by lattice coding is within $\frac{\log 3}{2}$ bits of the total exchange capacity and this gap decays to $0$ as $LP$ goes to infinity, i.e., lattice coding *achieves* exchange capacity if either the number of users or the power constraint of the users goes to infinity.

In Fig. 6, we illustrate the upper bound and the achievable total exchange rates for the pairwise data exchange model with $L = 8$ pairs as functions of $P$, while $P_r = 2LP$. Similar observations

$X_{L1}$
$X_{LK}$
$Y_{11}$
$Y_{1K}$
$Y_{L1}$
$Y_{LK}$
$W_{11}$
$W_{1K}$
$W_{L1}$
$W_{LK}$
$(\hat{W}_{11}^1,...,\hat{W}_{1K}^1)$
$(\hat{W}_{11}^K,...,\hat{W}_{1K}^K)$
$(\hat{W}_{L1}^1,...,\hat{W}_{LK}^1)$
$(\hat{W}_{L1}^K,...,\hat{W}_{LK}^K)$
Multi-way
Relay Channel
Cluster 1
Cluster $L$

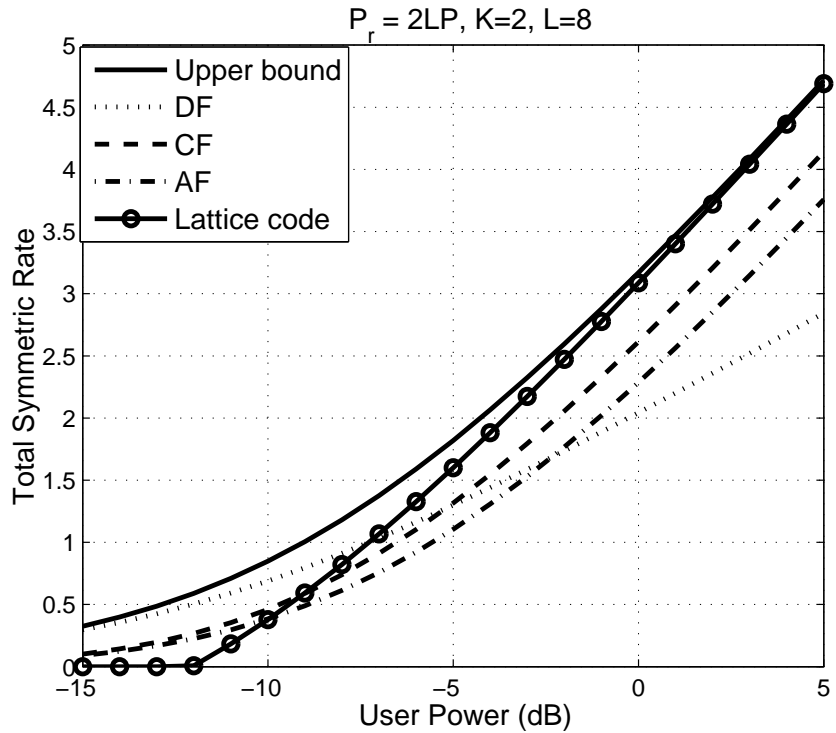

Fig. 6. Exchange capacity upper bound and achievable rates versus power $P$ for the pairwise data exchange model.

as in Section V-A apply for DF, CF and AF schemes. As the power constraint $P$ increases, lattice coding quickly outperforms other schemes and gets very close to the upper bound. In Fig. 7 we plot the total exchange rates with respect to the number of pairs in the system for $P = -5$ dB and $P_r = 2LP$. We can see that, similar to the behavior seen in Fig. 6, lattice coding improves as the number of pairs increases and gets very close to the upper bound, while CF and AF follow the upper bound within a finite bit gap uniform over the power constraints.

## VI. CONCLUSION

We have considered the Gaussian multi-way relay channel in which multiple clusters of users communicate simultaneously over a single relay terminal (no cross-reception between the users), and the users in each cluster want to exchange information among themselves. We have characterized the achievable rate region with AF, DF and CF schemes. When each cluster is composed of two users, we have characterized the rate region achievable by nested lattice coding as well. Specializing our results to the case of exchange rate points over symmetric networks,

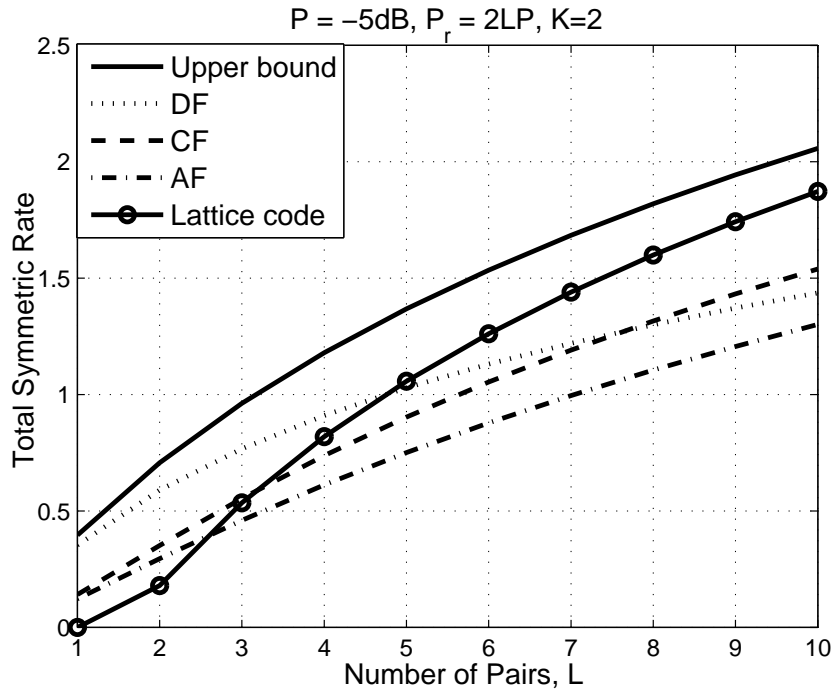

$X_{LK}$
$Y_{11}$
$Y_{1K}$
$Y_{L1}$
$Y_{LK}$
$W_{11}$
$W_{1K}$
$W_{L1}$
$W_{LK}$
$(\hat{W}_{11}^1,...,\hat{W}_{1K}^1)$
$(\hat{W}_{11}^K,...,\hat{W}_{1K}^K)$
$(\hat{W}_{L1}^1,...,\hat{W}_{LK}^1)$
$(\hat{W}_{L1}^K,...,\hat{W}_{LK}^K)$
Multi-way
Relay Channel
Cluster 1
Cluster $L$

Fig. 7. Exchange capacity upper bound and achievable rates versus the number of pairs in the system for the pairwise data exchange model.

we have shown that the CF scheme achieves exchange rates within a constant bit offset from the exchange capacity, while this constant gap is independent of the number of clusters and the power constraints of the nodes. The gap between the total exchange rates achieved by CF and AF schemes is also shown to be below a certain finite number of bits. Finally, we have shown that the nested lattice codes achieve rates within a finite bit gap of the exchange capacity for the case of multiple clusters with two users each, and that lattice coding outperforms all other schemes in this setup.

These results point to the fact that the additional decoding requirement at the relay node, imposed in the case of DF relaying, might be limiting in terms of the achievable exchange rates, and relaxing this requirement might lead to rates that are very close to the capacity in certain scenarios. While the decoding requirement is completely removed in the case of AF and CF relaying, it is relaxed in the case of lattice coding. It is an interesting research direction to explore other decoding requirements at the relay terminal with structured codes, that will be helpful in the case of clusters with multiple users. Our results provide design insights, in particular towards the relaying techniques to be employed and how close their performance is to the ultimate capacity bounds for this multi-way cooperative communication model.

| Block | 1 | 2 | $\cdots$ | $B$ | $B+1$ |
|---|---|---|---|---|---|
| $T_1$ | $x_1^n(w_{1,1})$ | $x_1^n(w_{1,2})$ | $\cdots$ | $x_1^n(w_{1,b})$ | $x_1^n(1)$ |
| $\vdots$ | | | | | |
| $T_K$ | $x_K^n(w_{K,1})$ | $x_K^n(w_{K,2})$ | $\cdots$ | $x_K^n(w_{K,b})$ | $x_K^n(1)$ |
| $Y_r$ | $\hat{y}_r(s_1)$ | $\hat{y}_r(s_2)$ | $\cdots$ | $\hat{y}_r(s_b)$ | |
| $X_r$ | $x_r(1)$ | $x_r(s_1)$ | $\cdots$ | $x_r(s_{b-1})$ | $x_r(s_b)$ |

Fig. 8. Illustration of the CF coding scheme.

## APPENDIX A
## PROOF OF THEOREM A

For simplicity of notation, we consider the case with a constant $Q$. The achievability for general time-sharing random variable can be obtained by using the classical arguments [22].

A block Markov encoding structure is considered, in which the messages are coded into $B$ blocks, and are transmitted over $B+1$ channel blocks. The relay forwards the information relating to each message block over the next channel block. The relay is kept silent in the first channel block, while the transmitters are silent in the last one. The receivers decode the messages from the relay's transmission right after each block. Since there is no coherent combining, transmitters send only new messages over each channel block, and thus sequential decoding over each block is sufficient.

*Codebook generation:* Fix $p(x_1) \cdots p(x_K) p(x_r) p(\hat{y}_r | y_r)$. The random codebook at user $i$ is generated i.i.d. from the distribution $\prod_{j=1}^{n} p(x_{i,j})$ for each message $w_i \in [1, 2^{nR_i}]$. We also generate $2^{nR_Q}$ quantization codewords i.i.d. according to $\prod_{j=1}^{n} p(\hat{y}_{r,i})$, and for each of these we generate one relay codeword i.i.d. with $\prod_{j=1}^{n} p(x_{r,j})$. We enumerate these codewords as $\hat{y}_r^n(w)$ and $x_r^n(w)$, respectively, for $w \in \{1, \ldots, 2^{nR_Q}\}$.

*Encoding:* See Figure 8 for an illustration of the encoding scheme over the channel blocks. Transmitter $T_i$ transmits the codeword $x_i^n(w_{i,b})$ at channel block $b = 1, \ldots, B$. All users transmit the codewords corresponding to message index 1 at the last channel block. The relay, upon receiving $y_r^n(b)$, looks for an index $s_b$ such that the corresponding codeword $\hat{Y}_r^n(s_b)$ is jointly typical with $y_r^n(b)$, i.e., $(y_r^n(b), \hat{Y}_r^n(s_b)) \in T^n_{[Y_r \hat{Y}_r]_\delta}$[1]. If no, or more than one such $s_b$ is found, it sets $s_b = 1$. Then the relay transmits $X_r^n(s_b)$ in the channel block $b+1$.

*Decoding:* Upon receiving $y_i^n(b)$, the user $i$ looks for the set of messages indices

$$(\hat{w}^i_{1,b-1}, \ldots, \hat{w}^i_{i-1,b-1}, w_{i,b-1}, \hat{w}^i_{i+1,b-1}, \ldots, \hat{w}^i_{K,b-1})$$

such that there exists an index $s \in [1, 2^{nR_Q}]$ for which

$$(X_1^n(\hat{w}^i_{1,b-1}), \ldots, X_{i-1}^n(\hat{w}^i_{i-1,b-1}), x_i^n(w_{i,b-1}), X_{i+1}^n(\hat{w}^i_{i+1,b-1}), \ldots, X_K^n(\hat{w}^i_{K,b-1}), \hat{Y}_r^n(s)) \in T^n_{[X_1\ldots\ldots X_K \hat{Y}_r]_\delta},$$

and

$$(X_r^n(s), y_i^n(b)) \in T^n_{[X_r Y_i]_\delta}$$

are simultaneously satisfied. If no, or more than one such set of message indices are found, then we set $\hat{w}^i_{1,b-1} = \cdots = \hat{w}^i_{i-1,b-1} = \hat{w}^i_{i+1,b-1} = \cdots = \hat{w}^i_{K,b-1} = 1$.

*Error Analysis:* Note that the error probability at each channel block is independent from the others. Hence, we will consider the error probability for each channel block separately as the total error probability will be bounded by the sum. For simplicity we will drop the channel block indices in the variables. Let $W_1, \ldots, W_K$ denote the messages of the users, and $S$ denote the quantization index chosen by the relay.

We define the following error events.

$$\varepsilon_1 \triangleq \left\{ (y_r^n, \hat{Y}_r^n(s)) \notin T^n_{[Y_r \hat{Y}_r]_\delta} \text{ for any } s \in [1, 2^{nR_Q}] \right\},$$

and

$$\varepsilon_2^i(w_1, \ldots, w_{i-1}, w_{i+1}, \ldots, w_K) \triangleq$$
$$\{(X_1^n(w_1), \ldots, X_{i-1}^n(w_{i-1}), x_i^n(w_i), X_{i+1}^n(w_{i+1}), \ldots, X_K^n(w_K), \hat{Y}_r^n(s)) \in T^n_{[X_1\ldots\ldots X_K \hat{Y}_r]_\delta}, \text{ and }$$
$$(X_r^n(s), y_i^n) \in T^n_{[X_r Y_i]_\delta} \text{ for some } s \in [1, 2^{nR_Q}]\}, \tag{31}$$

Assuming, without loss of generality, that $W_1 = \cdots = W_K = 1$ and $S = 1$, the error probability can be upper bounded by

$$\mathrm{P}(\varepsilon_1) + \mathrm{P}(\cup_{i=1}^K \varepsilon_1^c \cap \varepsilon_2^{i,c}(\mathbb{1})) + \mathrm{P}(\cup_{i=1}^K \cup_{\mathbf{w} \neq \mathbb{1}} \varepsilon_2^{i,c}(\mathbf{w})),$$

---

[1]The set of $\delta$- typical $n$-tuples according to $P_X$ is denoted by $T^n_{[X]_\delta}$.

where $\varepsilon^c$ is the complement of the event $\varepsilon$, $\mathbf{w} \in \mathbb{Z}^{K-1}$ with $\mathbf{w}_i \in [1, 2^{nR_i}]$ and $\mathbb{1}$ is the length-$(K-1)$ vector of 1s. We can further upper bound this by

$$P(\varepsilon_1) + \sum_{i=1}^{K} P(\varepsilon_1^c \cap \varepsilon_2^{i,c}(\mathbb{1})) + \sum_{i=1}^{K} P(\cup_{\mathbf{w} \neq \mathbb{1}} \varepsilon_2^{i,c}(\mathbf{w})).$$

Note that, as $n \to \infty$, $P(\varepsilon_1) \to 0$ if $R_Q = I(Y_r; \hat{Y}_r) + \epsilon$, and $P(\varepsilon_1^c \cap \varepsilon_2^{i,c}(1, \ldots, 1)) \to 0$ from the properties of the typical sets [22].

We can bound the last error term as follows.

$$P(\cup_{\mathbf{w} \neq \mathbb{1}} \varepsilon_2^{i,c}(\mathbf{w})) = \sum_{\mathcal{S} \subset \mathcal{I}_K \setminus \{i\}} \sum_{\mathbf{w}: w_j = 1 \Leftrightarrow j \in \mathcal{S}} P(\cup_{\mathbf{w} \neq \mathbb{1}} \varepsilon_2^{i,c}(\mathbf{w}|s=1)) + P(\cup_{\mathbf{w} \neq \mathbb{1}} \varepsilon_2^{i,c}(\mathbf{w}|s \neq 1)) \quad (32)$$

$$= \sum_{\mathcal{S} \subset \mathcal{I}_K \setminus \{i\}} 2^{nR(\mathcal{S})} \cdot 2^{-n(I(X(\mathcal{S}); \hat{Y}_r | X(\mathcal{S}^c)) - \epsilon')}$$

$$+ 2^{nR(\mathcal{S})} \cdot 2^{nR_Q} \cdot 2^{-n(I(X^K; \hat{Y}_r) - \epsilon')} \cdot 2^{-n(I(X_r; Y_i) - \epsilon'')} \quad (33)$$

These two terms go to 0 as $n \to \infty$ if

$$R(\mathcal{S}) < \min\{I(X(\mathcal{S}); \hat{Y}_r | X(\mathcal{S}^c)), I(X^K; \hat{Y}_r) + I(X_r; Y_i) - I(Y_r; \hat{Y}_r)\} - \bar{\epsilon} \quad (34)$$

$$= \min\{I(X(\mathcal{S}); \hat{Y}_r | X(\mathcal{S}^c)), I(X_r; Y_i) - I(Y_r; \hat{Y}_r | X^K)\} - \bar{\epsilon}, \quad (35)$$

for appropriately chosen positive $\epsilon, \epsilon', \epsilon''$ and $\bar{\epsilon}$.

Hence, the rates should satisfy the following set of inequalities

$$\sum_{k \in \mathcal{S}} R_k^{CF} \leq \min_{t \in \mathcal{S}^c} \min\{I(X(\mathcal{S}); \hat{Y}_r | X(\mathcal{S}^c), Q), I(X_r; Y_t | Q) - I(Y_r; \hat{Y}_r | X^K, Q)\}, \quad (36)$$

for all $\mathcal{S} \subset \mathcal{I}_K$, which is equivalent to the form given in the theorem.